\documentclass[twocolumn,amsmath,amssymb,prl]{revtex4}
\usepackage{mathrsfs}
\usepackage{graphicx,epsf,subfigure}

\begin{document}

\author{
J\'er\'emy Hure
}
\author{
Beno\^it Roman
}
\author{Jos\'e Bico
}

\title{Wrapping an adhesive sphere with a sheet}

\affiliation{Physique et M\'ecanique des Milieux H\'et\'erog\`enes, CNRS UMR 7636, UPMC \& Univ. Paris Diderot, ESPCI-ParisTech, 10 rue Vauquelin, 75231 Paris Cedex 05, France.}

\begin{abstract}
We study the adhesion of an elastic sheet on a rigid spherical substrate. Gauss'\textit{Theorema Egregium} shows that this operation necessarily generates metric distortions (i.e. stretching) as well as bending. As a result, a large variety of contact patterns ranging from simple disks to complex branched shapes are observed as a function of both geometrical and material properties.
We describe these different morphologies as a function of two non-dimensional parameters comparing respectively bending and stretching energies to adhesion. A complete configuration diagram is finally proposed. 
\end{abstract}

\maketitle  

Different types of projections have been developed to map the earth, such as the Mercator projection \cite{mercator} widely used for navigation purposes. 
Cartographers creating these projections face the challenge to transform a sphere into a planar region.
However Gauss proved in his \textit{Theorema Egregium} that such an operation cannot preserve both areas and angles. 
Indeed the product of the principal curvatures is constant under local isometry \cite{struik}. 
In other words, Gauss' theorem states that it is impossible to flatten a tangerine peel without tearing it. 
As a consequence, the length scale on a Mercator conformal map (which does preserve the angles) depends on the latitude. 
Sailors searching for the shortest route to cross the oceans thus follow curved paths on such maps. 
From a technological point of view, covering a curved substrate with a flexible surface is however a common operation. 
For instance, placing a contact lens over an eye of a mismatched geometry induces stresses in lenses \cite{funken96} and wrapping a sphere with a flat paper generates wrinkles \cite{demaine2009}.
As a practical consequence, bandages dedicated to knuckles or nose are tailored into specific templates in order to provide a good adhesion on round body parts  \cite{patent2}.
Understanding the adhesion of vesicles on curved substrates is also crucial for some drug delivery applications \cite{das2008}.  
In the field of microtechnology, special processes for depositing thin films \cite{fries02} or components \cite{ko2008} on curved substrates have been developed especially to account for the geometrical constraints dictated by Gauss'\textit{Theorema Egregium}. 
New theoretical approaches have also been recently developed to account for the specific crystallographic properties of crystals lying on curved substrates \cite{vitelli2006}.
The contact between a graphene sheet and a corrugated soft substrate finally allows to estimate the adhesion energy and bending stiffness of the graphene sheet \cite{scharfenberg2010},  which leads to novel metrology techniques. 

\begin{figure}[!h]
\begin{minipage}[c]{0.45\linewidth}
\subfigure[]{\includegraphics[width=4.cm]{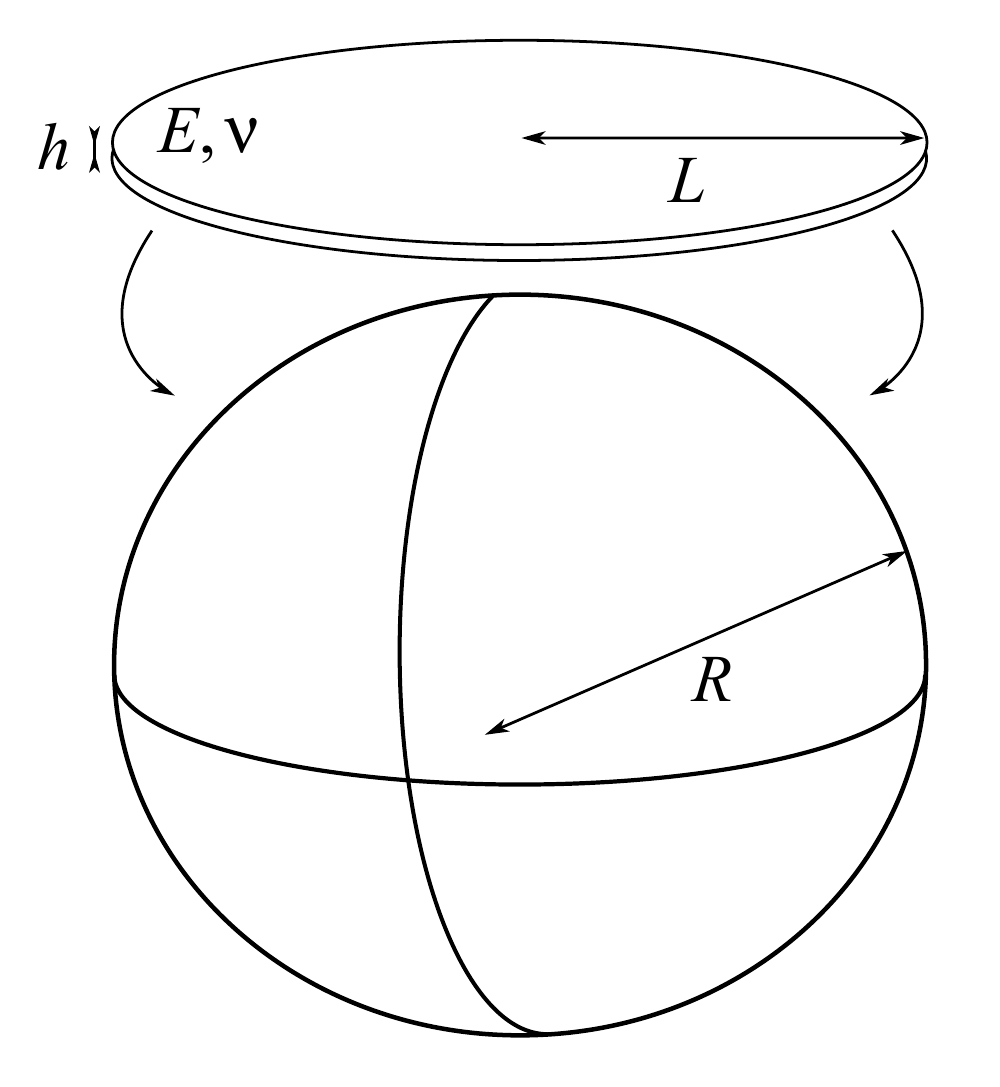}}
\end{minipage}
\hspace{0.5cm}
\begin{minipage}[c]{0.45\linewidth}
\subfigure[]{\includegraphics[width=4.cm]{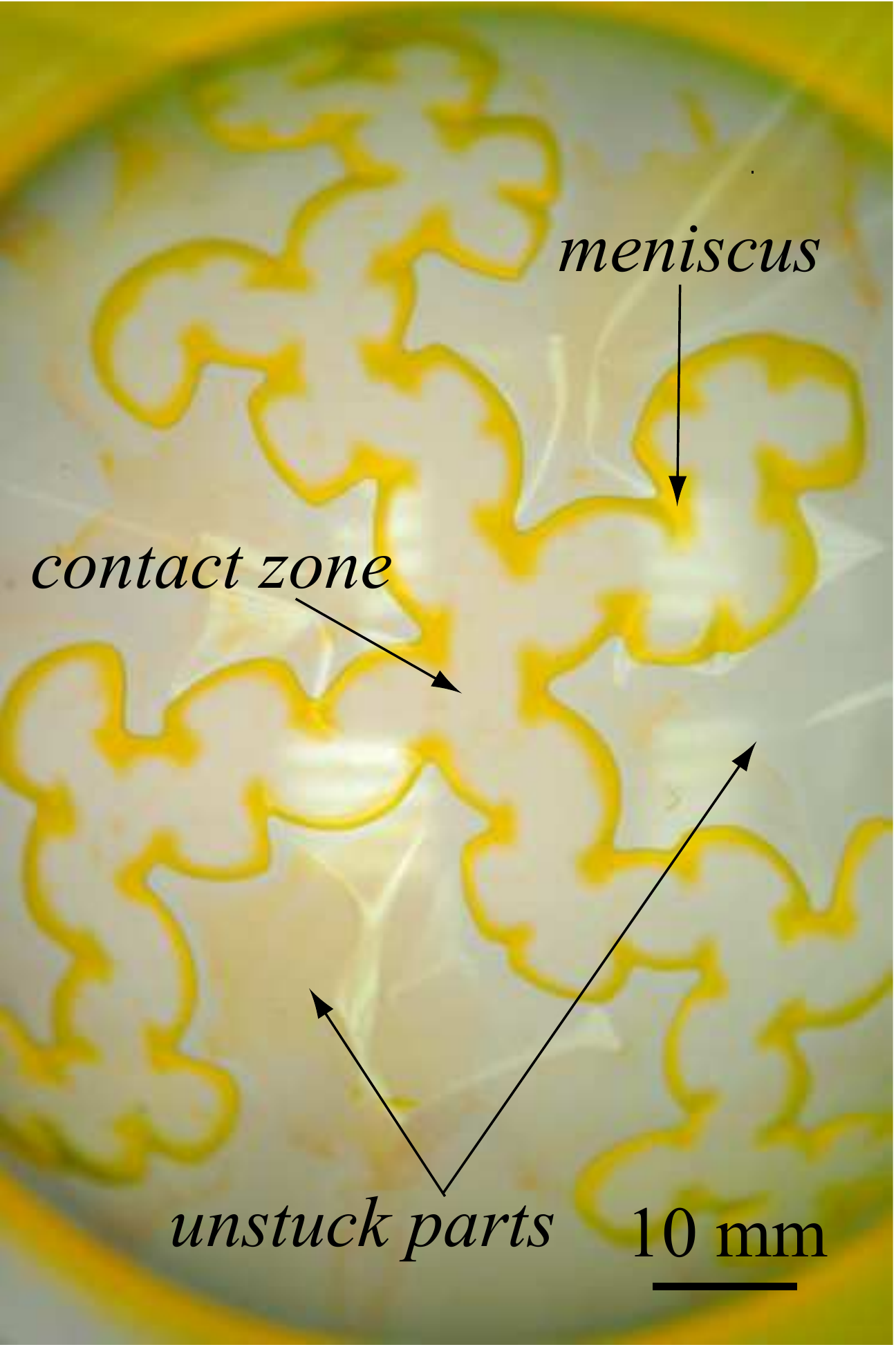}}
\end{minipage}
\vspace{-2mm}
\caption{(a) Experimental setup: an elastic plate of typical size $L$, Young's modulus $E$, Poisson's ratio $\nu$ and thickness $h$ is laid down a rigid sphere of radius $R$ preliminary coated with ethanol (surface tension $\gamma=22.4\, \mathrm{mN.m^{-1}}$). 
Ethanol totally wets both the plate and the sphere. 
(b) Typical experimental observation: ($E=2.6\,$GPa, $\nu=0.4$, $h=15\,\mu$m, $R=60\,$mm). 
In this example the region in contact with the sphere ({\it contact zone}) forms branched wavy patterns, while the {\it unstuck parts} of the sheet do not touch the sphere. 
A fluorescent dye allows to visualize the liquid meniscus that delimits both regions.
}
\label{fig1} 
\end{figure}

We propose to study, through model experiments, the reciprocal problem of the cartographer, \textit{i.e.} transforming a planar elastic sheet into a portion of sphere.
A thin film is deposited on a rigid spherical cap coated with a thin liquid layer (Fig.~\ref{fig1}a). 
Surface tension promotes the contact between the film and the sphere, which reduces the liquid/air interfacial energy at the cost of bending and stretching energies in the film. 
Experiments were conducted with polypropylene films (Innovia films) of four different thicknesses $h=15,30,50$ and $90\,\mu$m. 
The Young's modulus and the Poisson's ratio of the polymer are $E=2.6\pm0.2\,$GPa and $\nu=0.4$, respectively.
Prior to experiments, rigid polystyrene or glass spheres of radius ranging from $25\,$mm to $500\,$mm were coated with a layer of  ethanol  of surface tension $\gamma=22.4\, \mathrm{mN.m^{-1}}$, which allows the sheets to adhere on the spheres (ethanol totally wets both the spheres and the films).  
Depending on the parameters of the system, different morphologies of the contact zone between the sphere and the film are observed, spanning from total contact to branched patterns involving zigzagging contact zones and large unstuck parts (Fig.~\ref{fig1}b). 
Before describing the complex case of a sheet we first consider the simplified situation of an axisymmetric portion of a sheet: a thin elastic annulus deposited on a sphere. 
We proceed by highlighting the relevant physical parameters involved when an elastic plate adheres to a rigid sphere. 
We then study the typical size of contact between the plate and the sphere, as well as the contact pattern.
We finally present a configuration diagram showing the pattern observed as a function of the relevant physical parameters.\\  

\begin{figure}[!h]
\begin{minipage}[c]{0.42\linewidth}
\subfigure[]{\includegraphics[width=3.8cm]{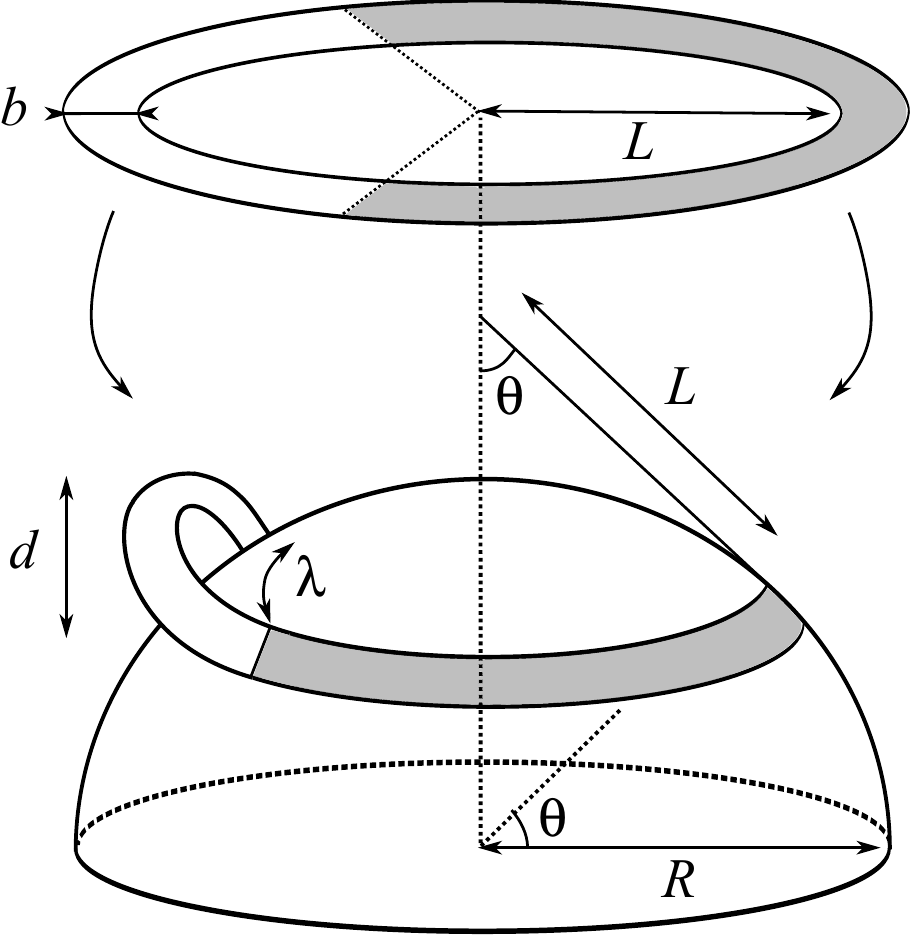}}
\end{minipage} 
\hspace{1cm}
\begin{minipage}[c]{0.42\linewidth}
\subfigure[]{\includegraphics[width=3.6cm]{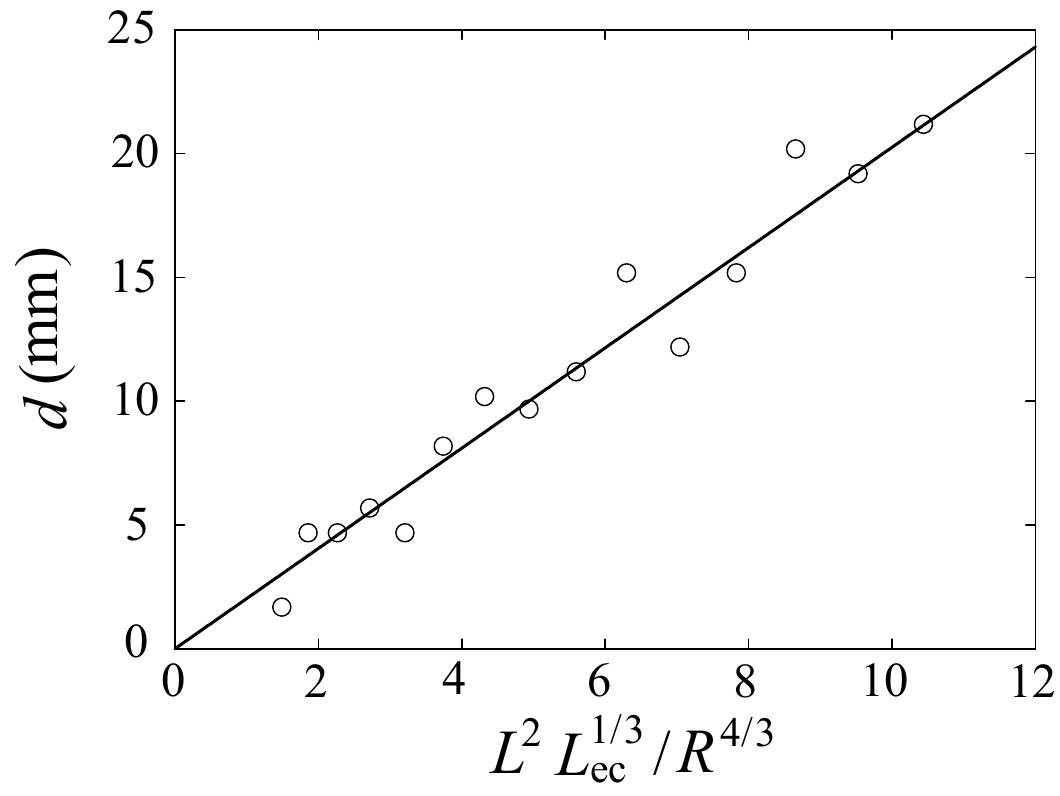}}
\subfigure[]{\includegraphics[width=3.6cm]{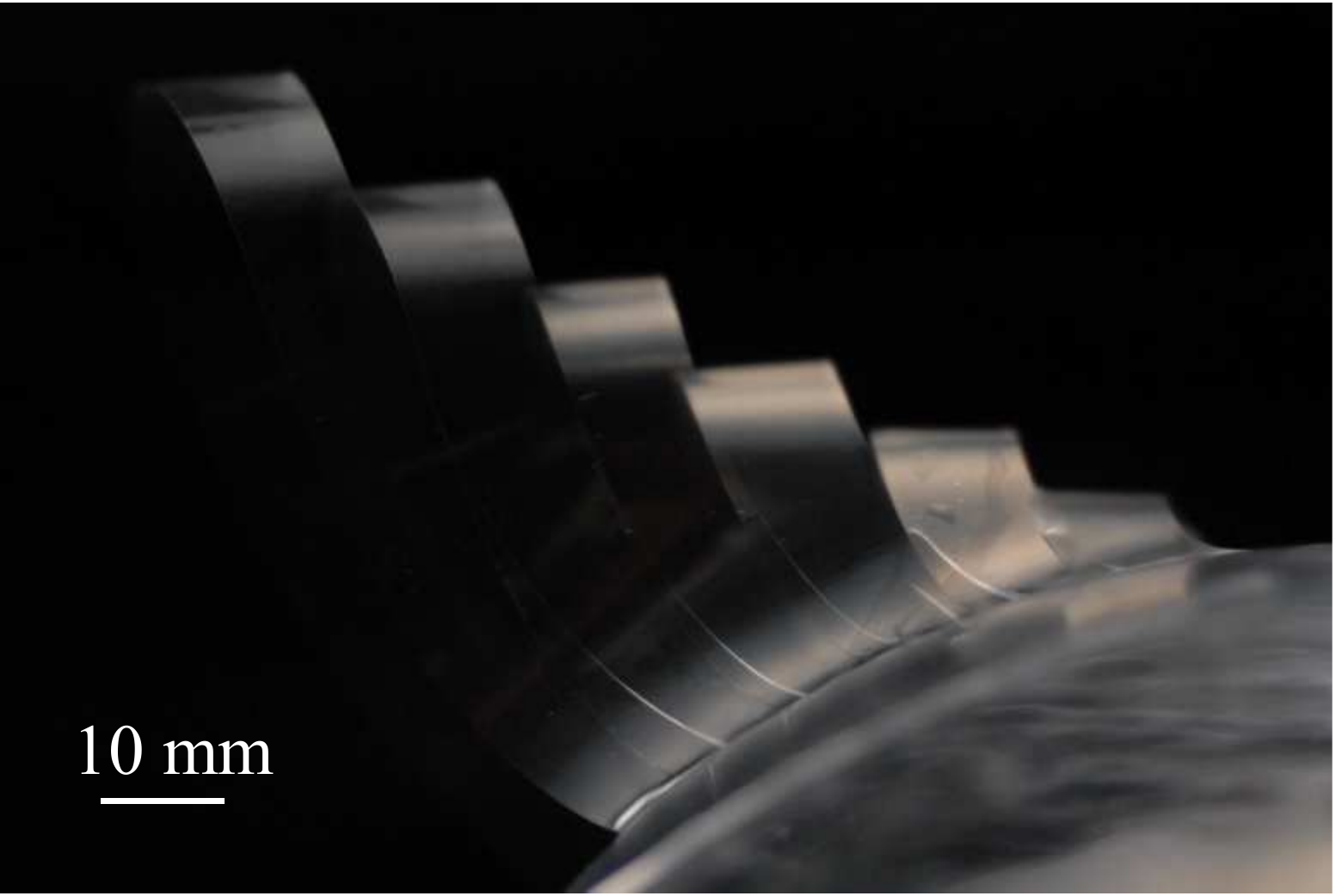}}
\end{minipage}
\vspace{-2mm}
\caption{(a) Experimental set-up: an elastic annulus of radius $L$ and width $b$ is deposited on a sphere coated with ethanol. (b) Height of the blister for a typical experiment ($R=100\,$mm, $L_{ec}=18\,$mm, $b=2\,$mm and $16.5\,\mathrm{mm}<L< 44.5\,\mathrm{mm}$). (c) Side view of the blisters obtained as elastic annuli of increasing radii are successively deposited on an adhesive sphere.}
\label{fig2} 
\end{figure}

Consider the simplified case of a flat elastic annulus delimited by concentric circles of respective radii  $L$ and $L+b$ (with $b \ll L$) deposited at the surface of an adhesive sphere of radius $R$  (Fig.~\ref{fig2}a). 
As observed in Fig.~\ref{fig2}c, only a limited portion of the annulus is in contact with the sphere, while the remaining part forms a unique blister of  height $d$ and width $\lambda$. 
We consider first the portion  of the annulus (in grey in Fig.~\ref{fig2}a) in contact with the sphere. The corresponding angular sector can be bent into a cone with half angle $\theta$ as shown in Fig.~\ref{fig2}a, with all lengths being conserved in this isometric operation. Only one angle $\theta$ allows to lay the annulus arc tangentially to the sphere of radius $R$, which is equivalent to contact in the limit $b \ll L$. This geometric condition sets the latitude $\theta$ and can be formulated in terms of geodesic curvature: 
to avoid local change of external and internal perimeters, the geodesic curvature $\kappa_g = \tan{\theta}/R$ of the annulus has to be equal to its initial planar curvature $1/L$, yielding $ \tan{\theta} = R/L$ \cite{struik}. 
However this geometrical constraint imposes a global excess of perimeter length for the annulus, $\Delta l = 2 \pi (L - R\cos \theta)$.
Within the limit of narrow annuli ($L\ll R$) this excess length scales as $\Delta l \sim L^3/R^2$.
The height and width of one dimensional blisters are dictated by a balance between adhesion and bending energies \cite{kendall76, vella09} and  leads to $d \sim \Delta l^{2/3} L_{ec}^{1/3} \sim L^2 L_{ec}^{1/3}/R^{4/3}$, where $L_{ec}=\sqrt{B/\gamma}$ is referred to as the \textit{elasto-capillary} length \cite{roman2010}, with  $B=Eh^3/[12(1-\nu^2)]$ corresponding to the bending modulus of the plate.
Experiments agree well with the prediction for the height of the blister (Fig.~\ref{fig2}b) confirming the above description where stretching energy has been neglected. A disk of radius $L$ can be seen as a collection of annuli, and it is tempting to consider the union of optimal (stretch-free) shapes for each annulus. 
However this solution involves at least radial compression, since the disk radius along the sphere $R(\pi/2-\theta)$ would be smaller than the initial $L=R\tan(\pi/2-\theta),$ leading to a typical radial strain $(L/R)^2.$ 
The corresponding stretching energy has thus to be taken into account in the description of the adhesion of the plain sheet.

We consider now the initial case of a disk of radius $L$ deposited on a sphere of radius $R$ coated with a wetting liquid. We restrict ourselves to the limit where the volume of liquid goes to zero, which is equivalent to consider dry adhesion without friction.
Mapping the sphere with the disk requires to bend the initially flat sheet, which involves a bending energy on the order of $\mathcal{E}_b \sim (B/R^2)L^2$ \cite{timoshenko}, the corresponding decrease in adhesion energy being $\mathcal{E}_{\gamma} \sim \gamma L^2$. Bending is thus promoted if $R$ is large in comparison to $L_{ec}$, which leads to a dimensionless parameter $R/L_{ec}$. Bending energy being predominant in the plate for small deflections, \textit{i.e.} for small contact areas, contact between the sphere and the plate is thus expected only for $R > L_{ec}$.

In addition to bending, stretching is also involved as previously illustrated with the annuli. In order to estimate the strain involved when the disk is forced to match the sphere, we assume that each perimeter of the plate remains of constant length. 
The variation in length in the radial direction is thus on the order of $\Delta l \sim L^3/R^2$, which corresponds to the typical strain  $\epsilon \sim \Delta l/L \sim (L/R)^2$ (this strain can also be quantitatively derived from classical F\"oppl-von K\'arm\'an equations \cite{landau, majidi2008}) and to the energy $\mathcal{E}_s \sim Eh L^2 \epsilon^2 \sim EhL^6/R^4$.
The balance of this stretching energy with adhesion leads to the dimensionless ratio  $L/\xi$, with  $\xi = R(\gamma/Eh)^{1/4}$. Within the limit $R \gg L_{ec}$, bending is negligible compared to adhesion. The extension of the contact zone should then be dictated by an equilibrium between stretching and adhesion energies, and is thus expected to scale as $\xi$.

\begin{figure}[!h]
\begin{minipage}[b]{0.5\textwidth}
\subfigure[]{\includegraphics[width=7.2cm]{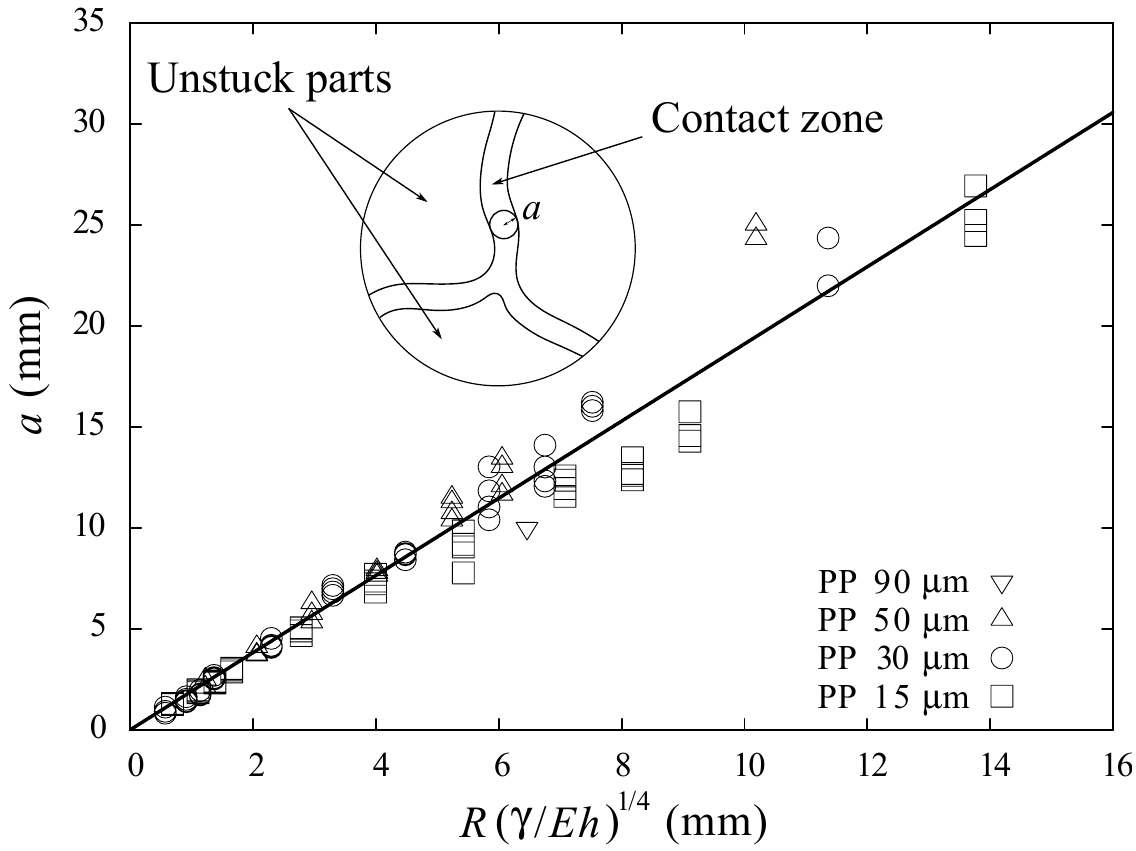}}
\end{minipage}
\begin{minipage}[b]{0.5\textwidth}
\subfigure[]{\includegraphics[height=3.cm]{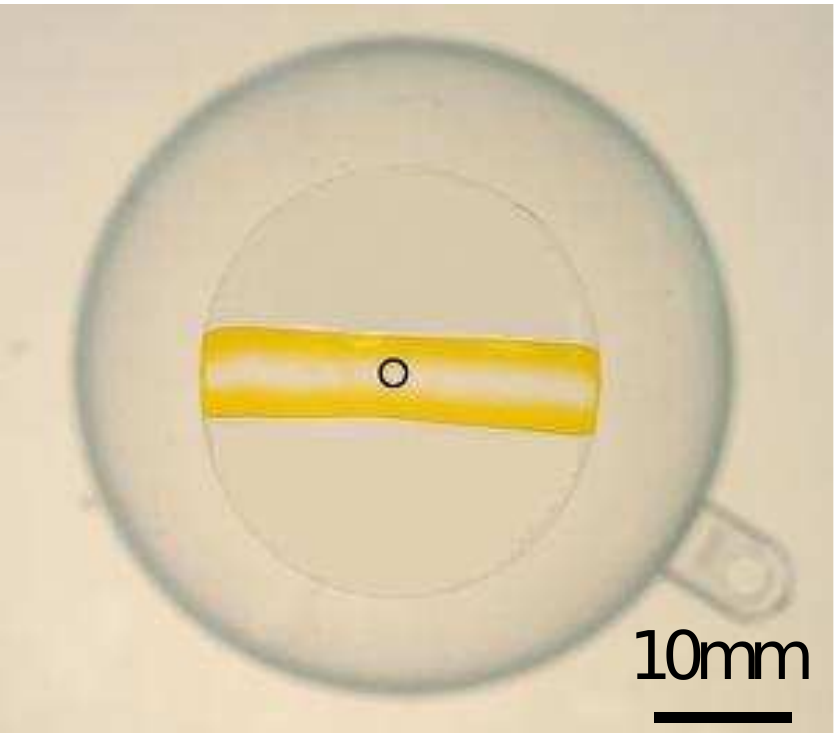}}
\subfigure[]{\includegraphics[height=3.cm]{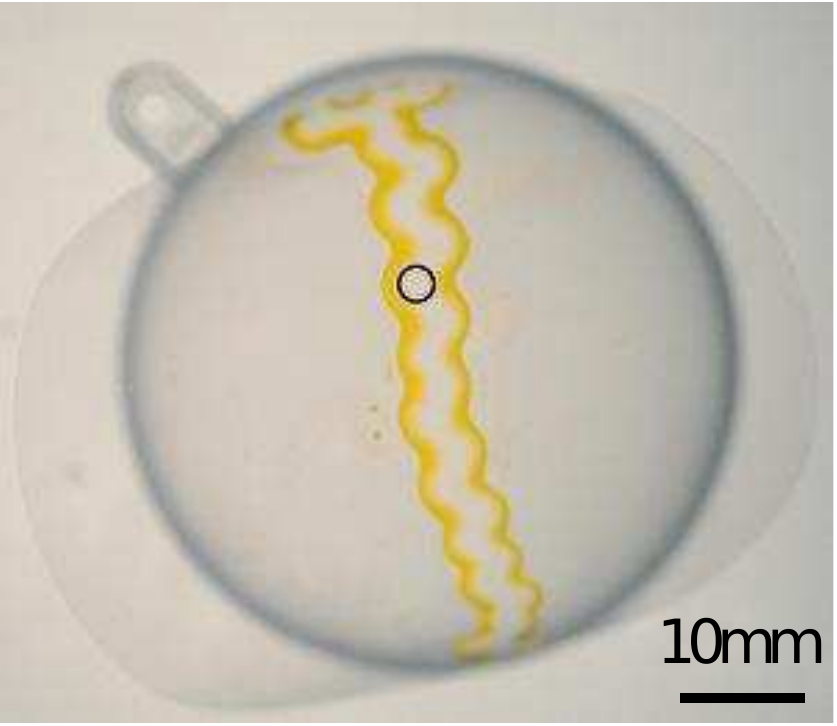}}\\
\subfigure[]{\includegraphics[height=3.cm]{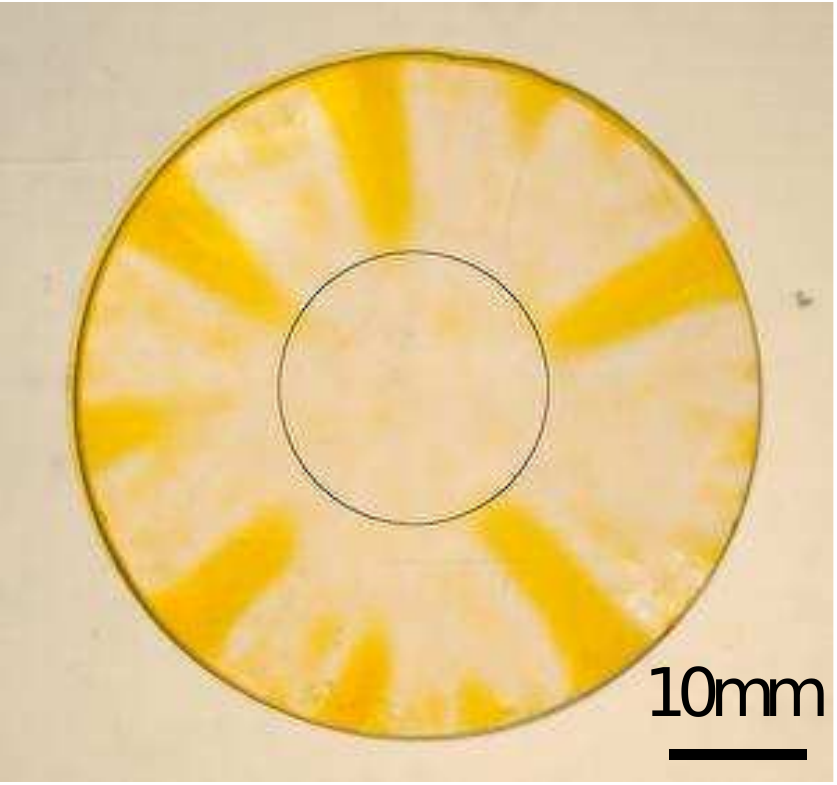}}
\subfigure[]{\includegraphics[height=3.cm]{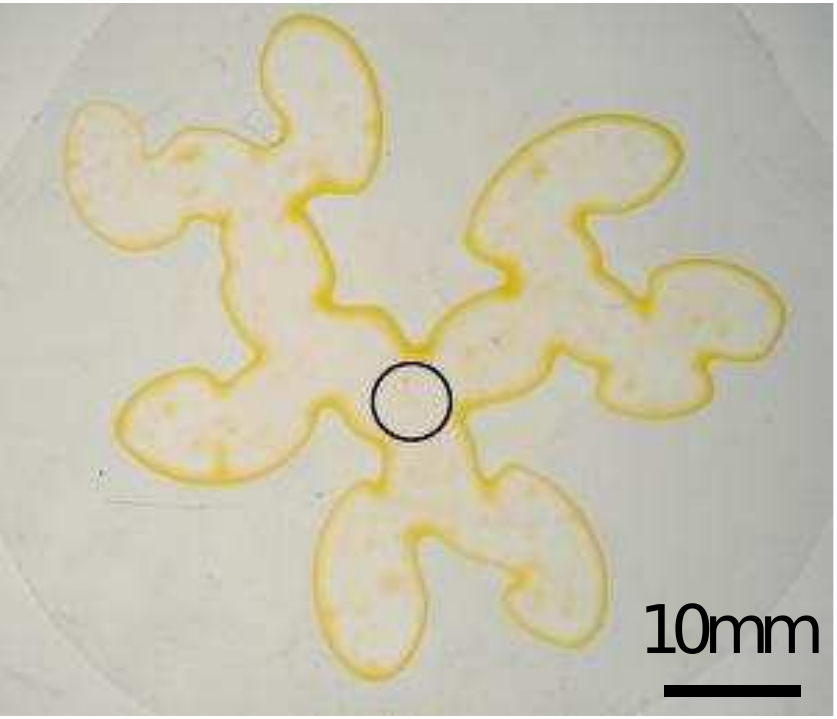}}
\end{minipage}
\vspace{-2mm}
\caption{(a) Size of the contact zone $a$, defined as the radius of the largest disk inscribed in the contact zone (insert), as a function of $R(\gamma/Eh)^{1/4}$ for four different thicknesses and radii of sphere spanning $25\,$mm to $500\,$mm. 
Solid line: linear fit $a =  1.9R(\gamma/Eh)^{1/4}$. 
(b) $R= 25\,$mm, $E=2.8\,$GPa, $h=30\,$$\mu$m. 
(c) $R= 25\,$mm, $E=2.6\,$GPa, $h=15\,$$\mu$m. 
(d) $R= 197\,$mm, $E=2.6\,$GPa, $h=15\,$$\mu$m. 
(e) $R= 50\,$mm, $E=2.6\,$GPa, $h=15\,$$\mu$m. }
\label{fig3} 
\end{figure}

We measured quantitatively  the size of the contact zone for the different patterns obtained in experiments in the regime $R \gg L_{ec}$. 
We define this size $a$ as the radius of the largest disk inscribed in the contact zone (insert in Fig.~\ref{fig3}a and black circles in Fig.~\ref{fig3}b-e). 
As expected, $a$ is found proportional to $\xi$ with a prefactor 1.9 (Fig.~\ref{fig3}a). 
Discrepancies may be attributed to boundary effects, that can locally change the stretching energy.
Moreover bending energy also tends to decrease the extension of the contact zone when $R/L_{ec}$ is close to unity. 
The maximum size of complete contact $a_{max}$ of a plate on a sphere can be more precisely written as $a_{max}=[(\alpha \gamma R^4/Eh)- \beta h^2 R^2]^{1/4}$, where the constants $\alpha$ and $\beta$ depend on the geometry of the plate. In the case of a disk, these constants are $\alpha = 256$ and $\beta = 32/3(1-\nu)$ (with $\mathcal{E}_{\gamma}=2\gamma S$) as demonstrated by Majidi and Fearing \cite{majidi2008}, while for a strip we found $\alpha = 36$ and $\beta = 3/2(1-\nu)$ \footnote{The slight difference with \cite{majidi2008} is due to the fact we do not make assumptions for the in-plane displacement field.}. The prefactor found experimentally for branched patterns is thus relatively close to the case of a strip (which would give a prefactor of 2.45).
A solution to ensure the complete adhesion of a plate of low bending rigidity on a sphere thus consists in cutting the plate into portions of widths smaller than $\xi$. 

\begin{figure}[!h]
\centering
\includegraphics[width=8.5cm]{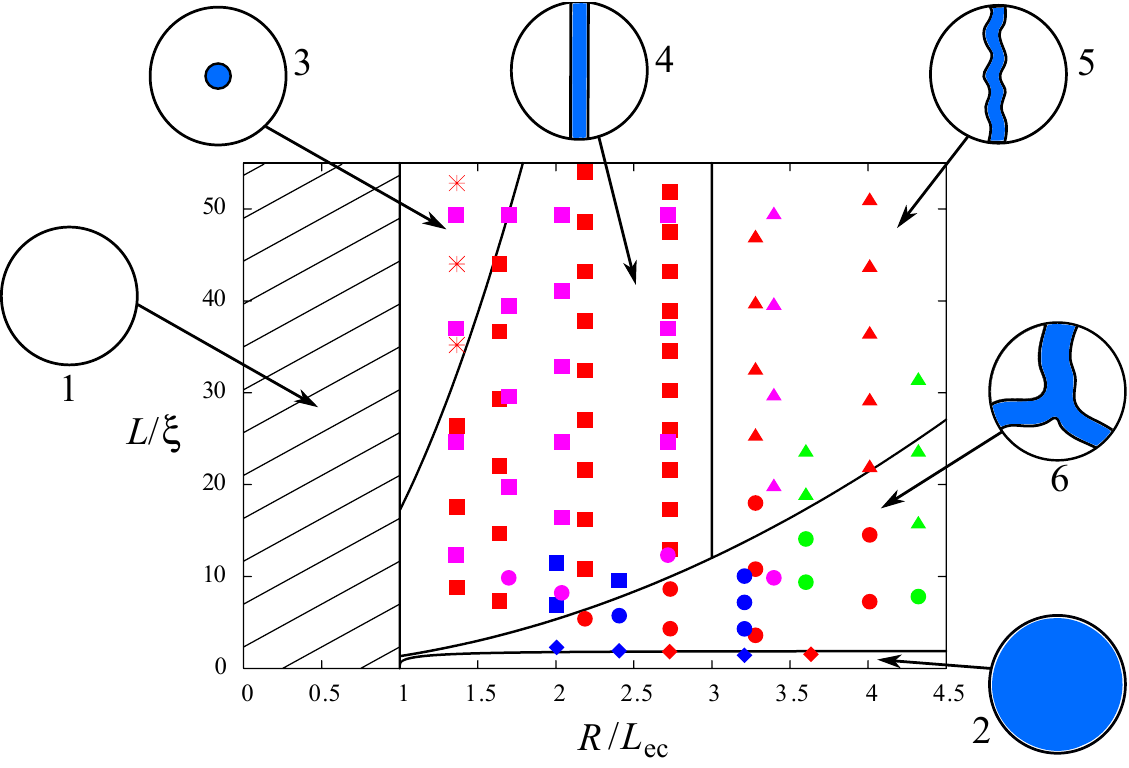}
\caption{Configuration diagram of the observed patterns of contact as a function of $R/L_{ec}$ and $L/\xi$, with $\xi=R(\gamma/Eh)^{1/4}$. Diamonds symbols correspond to complete contact (case 2), stars to local disk shaped contact (case 3), squares to straight strips (case 4), triangles to oscillating strips (case 5) and circles to branched patterns (case 6). Red symbols correspond to polypropylene ($E=2.6\,$GPa, $\nu=0.4$), blue ones to polyethylene ($E=170\,$MPa, $\nu=0.4$), green to natural rubber ($E=1.3\,$MPa, $\nu=0.5$) and pink to steel ($E=212\,$GPa, $\nu=0.3$).}
\label{fig4} 
\end{figure}

We now describe the geometry of the adhesion patterns as a function of the two parameters $R/L_{ec}$ and $L/\xi$.  As previously mentionned, when the ratio $R/L_{ec}<1$, the contact is limited to a point (case 1 in Fig.~\ref{fig4}).
Conversely, $R/L_{ec} \gg  1$ leads to a pattern of typical size $a \simeq 1.9\xi$. 
If the actual radius of the patch is smaller than $a$, a full coating should thus be observed (case 2 in Fig.~\ref{fig4}). 
The opposite situation is however richer: if we consider a fixed value for $L/\xi$ and progressively increase $R/L_{ec}$, patterns more and more complex are experimentally observed.
The lowest values of $R/L_{ec}$ only allow for a local bending of the sheet, which leads to a disk shaped contact zone (case 3 in Fig.~\ref{fig4}). 
Bending the whole sheet in one direction would indeed involve a greater bending energy on the order of $(B/R^2)L^2$, while the change in adhesion energy would be proportional to $\gamma L \xi$. 
We thus expect a strip-like adhesion pattern (case 4 in Fig.~\ref{fig4}) for $L/\xi < c_1 (R/L_{ec})^2$, where $c_1$ is a numerical prefactor. 
We found from our experiments $c_1= 19 \pm 3$. 
Wavy strips are observed for higher values of $R/L_{ec}$ (case 5 in Fig.~\ref{fig4}). 
Indeed, the contact between a strip on the plate and the sphere implies longitudinal stretching and compression along the contact edge. The transition to an oscillating pattern corresponds to an out-of-plane movement of the contact line, which release the in-plane compression. 
More quantitatively, this peculiar buckling instability occurs when the stretching energy density of the strip $Eh(\xi/R)^4\sim \gamma$ is of same order as the bending energy density induced by the out-of-plane bending of the sheet  $B/R^2$. This transition is thus expected when $R/L_{ec}$ becomes greater than a critical value of order 1.  
We found experimentally that this transition occurs for $R/L_{ec}>3\pm 0.3$. 
This instability is described in detail in a coming paper \cite{hure11}.
Branched patterns finally appear for higher values of $R/L_{ec}$ (case 6 in Fig.~\ref{fig4}). In this situation, the whole sheet is effectively bent in both directions. We expect the scaling laws for the bending energy, $(B/R^2)L^2$, and the adhesion energy, $\gamma L \xi$, to remain valid, but with larger prefactors than in the previous cases. 
Branched patterns should therefore develop for $L/\xi < c_2 (R/L_{ec})^2$, where $c_2$ is a numerical prefactor (lower than $c_1$). 
A fit with the experimental data indicates $c_2 = 1.3 \pm 0.3$. 
The different configurations observed for materials spanning five orders of magnitude of Young's modulus are depicted in Fig.~\ref{fig4}. 
The collapse of the experimental data confirms our scaling arguments for the transitions and the relevance of the pair of non-dimensional parameters $R/L_{ec}$ and $L/\xi$ to describe the adhesion patterns.\\

As can be observed on Fig.~\ref{fig1}b, the contact patterns are simply connected: branches never reconnect, and debonded areas always reach the edge of the sheet.
Indeed, consider a closed curve $\partial\cal{S}$ drawn on the sphere, along which the plate is in contact with the sphere (and therefore touches tangentially the sphere), but not on $\cal{S}$$ _p$, the plate surface. The corresponding surface on the sphere is noted $\cal{S}$$ _{s}$. Gauss-Bonnet theorem on the plate or on the spherical cap reads $\int_{\cal{S}} K + \int_{\partial\cal{S}} k_g = 2 \pi$, where $K$ and $k_g$ are the Gaussian curvature and the geodesic curvature, respectively \cite{struik}. 
This leads to $\int_{S_p} K  =\int_{S_{s}} K = \cal{S}$$ _{s}/R^2 $, as  $\partial\cal{S}$ belongs to both the plate and the sphere. 
The Gaussian curvature $K$ integrated on $\cal{S}$$ _p$ is thus finite and independent of the shape taken by the plate. 
According to the \textit{Theorema Egregium}, the plate bounded by $\partial\cal{S}$ is necessarily stretched, even if not in contact with the sphere.
The strain induced by the finite Gaussian curvature is given by $\Delta\epsilon  \sim K $ \cite{witten2007} and scales as $\epsilon \sim  \int_{\cal{S}_\mathrm{s}} K \sim \cal{S}$$ _{s}/R^2$ if the shape is characterized by a single typical dimension (elongated shapes are excluded). Stretching energy is thus as a first approximation independent of the actual shape of the plate. 
Since the decrease in adhesion energy is proportional to the contact surface, it is always energetically favorable to put in contact any region bounded by a closed contact line :  branched patterns cannot reconnect. \\

To summarize, a wide variety of adhesion patterns, ranging from full contact to branched shapes, are observed as an elastic sheet is laid down a rigid adhesive sphere. 
Due to the mismatch in Gaussian curvatures, wrapping the sphere involves finite stretching in the contact zone.
While a balance between stretching and adhesion energies provides the typical width of the zone $a\sim R(\gamma/Eh)^{1/4}$, the balance between bending and adhesion energies dictates the complexity of the pattern: simple disk, straight strip, oscillatory strip or branches. 
These different configurations can be predicted from two non-dimensional parameters $L/\xi \sim (Eh/\gamma)^{1/4}L/R$ and $R/L_{ec} \sim R\sqrt{\gamma/B}$.
Since surface forces become predominant at small scales \cite{roman2010}, we expect our results obtained through macroscopic experiments to be valid for micro- and nano-technologies. 
As an example, if a graphene monolayer
($E \simeq 1\,$TPa, $h \simeq 0.34\,$nm \cite{cranford09})
 is deposited on a silica bead of radius $R$ (with a Van der Waals adhesion energy of $W \simeq 500\,\mathrm{mJ.m^{-2}}$ \cite{li2009}) partial contact is expected for $R > 9 \AA$, with a contact width on the order of $0.2R$.  
Material properties (adhesion energy, mechanical stiffness) can finally be inferred from the analysis of the adhesion patterns, which may lead to a novel metrology technique relevant for thin films.\\

We thank Guillaume Batot and Dominique Vella for their help with preliminary experiments. This study was partially funded by the ANR project MecaWet.

\bibliographystyle{apsrev4-1}
\bibliography{spherebib}

\end{document}